\documentclass{aa} 

\usepackage{natbib}
\usepackage{graphicx}
\usepackage{txfonts}
\begin{document}

   \title{Possible detection of coronal mass ejections on late-type main-sequence stars in LAMOST medium-resolution spectra}


   \author{Hong-peng Lu
          \inst{1},
          Hui Tian\inst{1, 2},
          Li-yun Zhang\inst{3},
          Christoffer Karoff\inst{4,5},
          He-chao Chen\inst{1},
          Jian-rong Shi\inst{2},
          Zhen-yong Hou\inst{1},
          Ya-jie Chen\inst{1},
          Yu Xu\inst{1},
          Yu-chuan Wu\inst{1},
          Dong-tao Cao\inst{6},
          Jiang-tao Wang\inst{2}\\
          }

   \institute{School of Earth and Space Sciences, Peking University, Beijing 100871, People's Republic of China\\
              \email{huitian@pku.edu.cn}
         \and
             National Astronomical Observatories, Chinese Academy of Sciences, Beijing 100101, People's Republic of China
         \and
             College of Physics and Guizhou Provincial Key Laboratory of Public Big Data, Guizhou University, Guiyang 550025, People's Republic of China
          \and
             Stellar Astrophysics Centre, Department of Physics and Astronomy, Aarhus University, Ny Munkegade 120, DK-8000 Aarhus C, Denmark
          \and
             Department of Geoscience, Aarhus University, Høegh-Guldbergs Gade 2, DK-8000 Aarhus C, Denmark 
         \and
             Yunnan Observatories, Chinese Academy of Sciences, Kunming 650216, People's Republic of China
             }


 
  \abstract
   {Stellar coronal mass ejections (CMEs) are the primary driver of the exoplanetary space weather and they could affect the habitability of exoplanets. However, detections of possible stellar CME signatures are extremely rare. }
   {This work aims to detect stellar CMEs from time-domain spectra observed through the LAMOST Medium-Resolution Spectroscopic Survey (LAMOST-MRS). Our sample includes 1,379,408 LAMOST-MRS spectra of 226,194 late-type main-sequence stars ($\rm T_{eff} < 6000$ K, $\rm log [g/(cm\ s^{-2})] > 4.0$).}
   {We first identified stellar CME candidates by examining the asymmetries of H$\alpha$ line profiles, and then performed double Gaussian fitting for H$\alpha$ contrast profiles (differences between the CME spectra and reference spectra) of the CME candidates to analyze the temporal variation of the asymmetric components. }
   {Three stellar CME candidates were detected on three M dwarfs. The H$\alpha$ and Mg\,{\sc{i}} triplet lines (at 5168.94 Å, 5174.13 Å, 5185.10 Å) of candidate 1 all exhibit a blue-wing enhancement, and the corresponding Doppler shift of this enhancement shows a gradually increasing trend. The H$\alpha$ line also shows an obvious blue-wing enhancement in candidate 2. In candidate 3, the H$\alpha$ line shows an obvious red-wing enhancement, and the corresponding projected maximum velocity exceeds the surface escape velocity of the host star. The lower limit of the CME mass was estimated to be $\sim$$8 \times 10^{17}$ g to $4 \times 10^{18}$ g  for these three candidates.   }
   {}

   \keywords{techniques: spectroscopic --
                stars: late-type --
                Stars: solar-type --
                stars: activity --
                Sun: coronal mass ejections (CMEs)
               }

\titlerunning{Possible detection of CMEs on late-type main-sequence stars in LAMOST medium-resolution spectra}
\authorrunning{Lu et al.}
   \maketitle
%

\section{Introduction}

    Coronal mass ejections (CMEs), which are large-scale ejected structures consisting of plasma and magnetic field, are one of the most spectacular eruptive phenomena on stars \citep{1996SoPh..166..441H, 2000JGR...10523153F, 2004ApJ...602..422L, 2011LRSP....8....1C}. In star-exoplanet systems, frequently occurring CMEs may erode or even strip off the atmospheres of planets, rendering the planets uninhabitable \citep[e.g.,][]{2007AsBio...7..167K, 2007AsBio...7..185L, 2018arXiv180107333A, 2019LNP...955.....L}. In addition, large CMEs could generate high-energy particles, which may destroy ozone in the planetary atmospheres. This will allow a lot more UV photons to arrive at the planetary surfaces, affecting the habitability on planets \citep{2010AsBio..10..751S}. Moreover, CMEs could also contribute significantly to the loss of stellar angular momentum and mass during the long-term evolution of stars \citep{2007AsBio...7..167K, 2008SSRv..139..437Y, 2010ARA&A..48..241B, 2012ApJ...760....9A}.
   
   Solar CMEs have been frequently observed and intensively studied in the past decades. Depending on the solar activity level, normally there are 0.5--6 CME events per day on the Sun. The typical speed of solar CMEs is in the range of tens to thousands of kilometres per second, and the average mass of solar CMEs is about $4\times10^{14}$ g \citep{2010ASSP...19..289G, 2012LRSP....9....3W, 2017LRSP...14....5K, 2019LNP...955.....L}. Observations have shown that the maximum mass of solar CMEs is about $2 \times 10^{17}$ g, and the maximum kinetic energy of solar CMEs is about $1.2 \times 10^{33}$  erg \citep{2009EM&P..104..295G}.

  However, so far there have been only a few attempts of stellar CME detections. The main detection method is the Doppler-shift method, which searches for stellar CME signals by detecting the enhanced emission or absorption in the wings (or the nearby continuum) of spectral lines. With spatially resolved extreme-ultraviolet (EUV) spectroscopic observations,  \cite{2012ApJ...748..106T} reported obvious blue-wing enhancements in several spectral lines during a solar CME eruption. A CME-caused blue wing enhancement has also been detected in the Sun-as-a-star EUV spectra \citep{2022arXiv220411722X}. Inspired by these results, \cite{2022arXiv220403683Y} recently developed an analytical CME model and demonstrated that stellar CMEs may also be detected through EUV spectroscopy. By analyzing the X-ray spectra of a stellar flare on a G-type giant star (HR 9024), \cite{2019NatAs...3..742A} reported a blue-shifted component in the O\,{\sc viii}~18.97 Å line, which was ascribed to a possible CME associated with a flare. In the past, the Doppler-shift method were mostly applied to stellar spectra in the visible band. For example, through the blue-wing enhancement in the H$\gamma$ line, \cite{1990A&A...238..249H} discovered possibly the first stellar CME candidate with a maximum projected velocity of --5800 km s$^{-1}$ (here minus means blue shift) on AD Leo. With optical spectral observations of a solar-type star (EK Dra), a blue-shifted absorption component with a projected velocity of --510 km s$^{-1}$ in the H$\alpha$ line was detected, which may suggest the occurrence of a CME \citep{2022NatAs...6..241N}. A blue-wing enhancement in H$\alpha$ lasting for $\sim$60 minutes was reported by \cite{2021PASJ...73...44M}, which may result from a prominence eruption on YZ CMi. \cite{2020MNRAS.499.5047M} identified weak asymmetries of the Balmer lines in the spectra of EV Lac, which were attributed to an erupting filament. Based on the asymmetries of the H$\alpha$, H$\beta$ and H$\gamma$ lines on V374 Peg, \cite{2016A&A...590A..11V} detected a stellar CME candidate whose maximum projected velocity exceeds the stellar surface escape velocity. \cite{1997A&A...321..803G} identified a blue-shifted component with a projected velocity of about --600 km s$^{-1}$ in the H$\alpha$ line, which was interpreted as a CME on the observed T Tauri star. Similar Doppler-shifted emission or absorption features have also been reported and interpreted as stellar CMEs by several other authors \citep{1994A&A...285..489G, 2001ApJ...560..919B, 2004A&A...420.1079F, 2011A&A...536A..62L, 2019MNRAS.482..988C, 2020A&A...637A..13M}. Some authors have attempted to search for stellar CMEs using the Doppler-shift method, but no obvious CME signal (an obvious asymmetry in spectral lines) was found \citep{2014MNRAS.443..898L, 2017IAUS..328..198K, 2020MNRAS.493.4570L, 2021ApJ...916...92W}. \cite{2014MNRAS.443..898L} concluded that the signal to noise ratio (S/N), not the spectral resolution, is the most important factor that affects the application of the Doppler-shift method to search for stellar CMEs.

    In addition, there are studies trying to search for stellar CMEs from spectroscopic survey databases. \cite{2019A&A...623A..49V} used the virtual observatory data to search for stellar CMEs by analyzing the Balmer-line asymmetries of late-type stars. The typical projected velocity of their detected CME candidates is between 100 km s$^{-1}$ and 300 km s$^{-1}$, and the mass of the candidates ranges from $10^{12}$ kg to $10^{15}$ kg. An effort was made to hunt for CMEs from late-type main-sequence stars by analyzing the asymmetries of the Balmer lines in the SDSS spectra \citep{2021A&A...646A..34K}. Their results showed that the mass and projected maximum velocity of the CME candidates are $10^{16}$--$10^{18}$ g and 300--700 km s$^{-1}$, respectively.
   
   From solar observations, we know that energetic CMEs may generate type\,{\sc II} radio bursts. So a stellar type-II radio burst would be a strong evidence of stellar CMEs. Many authors have attempted to detect radio bursts on active stars (e.g. AD Leo, EV Lac). Although some radio bursts have been observed, there has been no unambiguous detection of type-II bursts \citep[e.g.,][]{1995A&AS..114..509A, 1997Ap&SS.257..131A, 1998A&AT...17..221A, 2009AIPC.1094..680L, 2012AASP....2..121B, 2016ApJ...830...24C, 2018ApJ...862..113C, 2021MNRAS.502.5438P}.  
    
   A new detection approach is through the stellar coronal dimming, which detects stellar CMEs by searching for a sudden dimming of the EUV or X-ray emission. \cite{2021NatAs...5..697V} verified the possibility of using the coronal dimming to detect stellar CMEs by treating the Sun as a star, and then used this method to detect multiple CME candidates on cool stars (e.g., AB Dor, AU Mic and Proxima Centauri). Besides, \cite{2017ApJ...850..191M} reported a CME candidate on the eclipsing binary Algol by the method of X-ray continuum absorption. Several cases of X-ray absorption in stellar observations may also be related to stellar CMEs \citep[e.g.,][]{1983ApJ...267..280H, 1996A&A...307..813O, 1998ApJ...503..894T, 2001A&A...375..196F, 2012MNRAS.419.1219P}. For other stellar CME detection methods and the detection history, please refer to \cite{2019ApJ...877..105M}.
   
   This work attempts to use the Doppler-shift method to search for possible stellar CMEs from late-type main-sequence stars in the LAMOST Medium-Resolution Spectroscopic Survey (LAMOST-MRS). The sample selection and CME detection method are presented in the second section. The third section presents a detailed analysis and discussion of the three identified CME candidates. The summary and future perspective are given in the final section.
   
\section{Data sample and CME detection method}
\subsection{LAMOST-MRS}
  
    The Large Sky Area Multi-Object Fiber Spectroscopic Telescope (LAMOST) \citep{2012RAA....12.1197C, 2012RAA....12..723Z, 2012RAA....12..735D, 2015RAA....15.1095L} is a four-meter reflective Schmidt telescope equipped with four thousand optical fibers. The field of view of the LAMOST is 20 square degrees. This telescope is located in Xinglong Station of National Astronomical Observatories, Chinese Academy of Sciences. From October 2011 to July 2018, the first phase of the LAMOST low-resolution spectroscopic survey was completed, collecting more than 10 million low-resolution stellar spectra (R$\sim$1800). Since October 2018, the second phase of the five-year spectroscopic surveys has been implemented, including the LAMOST low-resolution spectroscopic survey and the LAMOST medium-resolution spectroscopic survey (LAMOST-MRS), each of which accounted for half of the total observation time \citep{2020arXiv200507210L}. The LAMOST-MRS includes two modes: the time-domain survey and the non-time-domain survey. The time-domain survey repeatedly observes the targets on multiple observation nights. The LAMOST continuously exposes the same targets 3 to 8 times during each time-domain observation night. On a non-time-domain observation night, the LAMOST continuously observes the same targets 3 times. The typical exposure time is 1200 seconds, and the limiting magnitude in the G-band is about 15 mag. Therefore, the target observed by the LAMOST-MRS will have more than 3 continuously observed spectra.
    
    LAMOST-MRS raw data is processed by the LAMOST 2D pipeline, including standard dark and bias subtraction, flat-field correction, spectral extraction, sky subtraction and wavelength calibration \citep{2015RAA....15.1095L, 2020ApJS..251...15Z}. The LAMOST-MRS spectrum includes a blue arm (4950 -- 5350 Å) and a red arm (6300 -- 6800 Å), and the spectral resolving power R is $\sim$7500. The typical radial velocity accuracy of the LAMOST-MRS observations reaches 1 km s$^{-1}$ \citep{2019RAA....19...75L, 2019ApJS..244...27W}. The LAMOST-MRS also provides fundamental stellar atmospheric parameters including the stellar effective temperature ($\rm T_{eff}$) and the surface gravity ($\rm log\ g$), etc \citep{2020ApJ...891...23W}. From October 2018 to June 2021, the number of the LAMOST-MRS spectra in a single exposure exceeds 10 million. These spectra will be released through the LAMOST DR8\footnote{http://www.lamost.org/dr8/} and the DR9 v0\footnote{http://www.lamost.org/dr9/}.
    
 
  \begin{figure*}
   \centering
   \includegraphics[width=\hsize]{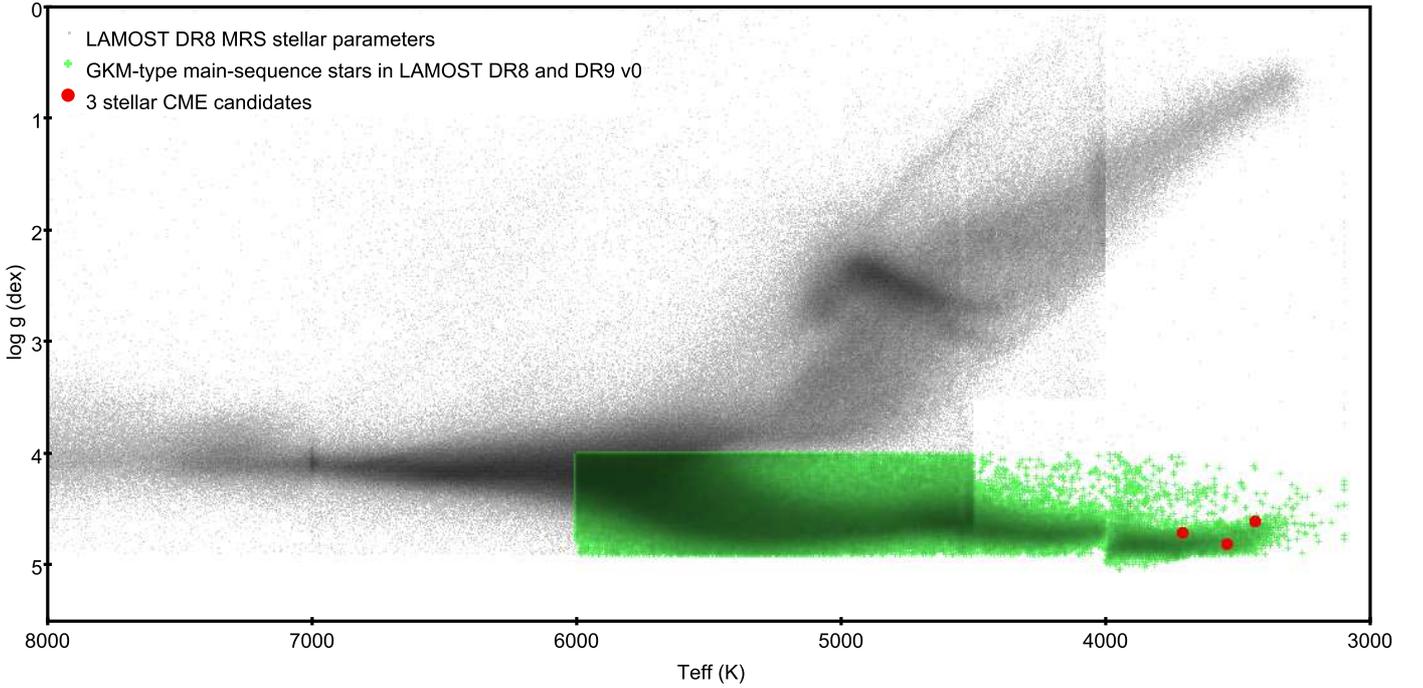}
      \caption{Location of the search sample (GKM-type main-sequence stars, $\rm T_{eff}$ < 6000 K, $\rm log [g/(cm\ s^{-2})] > 4.0$) on the Hertzsprung-Russell diagram. The stellar parameters corresponding to the gray squares are obtained from the LAMOST-MRS parameter catalog in the LAMOST DR8. The green crosses represent the sample for this study, their parameters come from the LAMOST-MRS parameter catalog in the LAMOST DR8 and DR9 v0. The red solid circles show the locations of the host stars for the three stellar CME candidates.    }
         \label{HR}
   \end{figure*}       

\subsection{Sample selection and Data processing}

    The purpose of this study is to search for stellar CMEs on late-type main-sequence stars in the LAMOST-MRS. We selected the late-type main-sequence stars based on the stellar effective temperature ($\rm T_{eff}$) and the surface gravity ($\rm log\ g$). The selection criteria are  $\rm T_{eff}$ < 6000 K and $\rm log [g/(cm\ s^{-2})] > 4.0$. In addition, the S/N of the red arm and the blue arm of each LAMOST-MRS spectrum is required to be greater than 5. The location of the final selected sample on the Hertzsprung-Russell diagram is shown in Fig. \ref{HR}. The numbers of different types of stars in the sample are summarized in Table \ref{table:1}. This sample includes 1,379,408 LAMOST-MRS spectra, which are from 226,194 late-type main-sequence stars, as shown in Table \ref{table:1}.
  
   Before detecting stellar CMEs, the LAMOST-MRS spectra need to be preprocessed. First, we normalized the red and blue arms of each LAMOST-MRS spectrum separately using the "laspec" toolkit \citep{2021ApJS..256...14Z}, and removed cosmic rays. Then the wavelength of each spectrum was corrected by using the radial velocity parameter (rv\_br0) in the LAMOST-MRS parameter catalog. These preprocessed spectra were then used to search for stellar CMEs.
    
 
  \begin{figure*}
   \centering
   \includegraphics[width=\hsize]{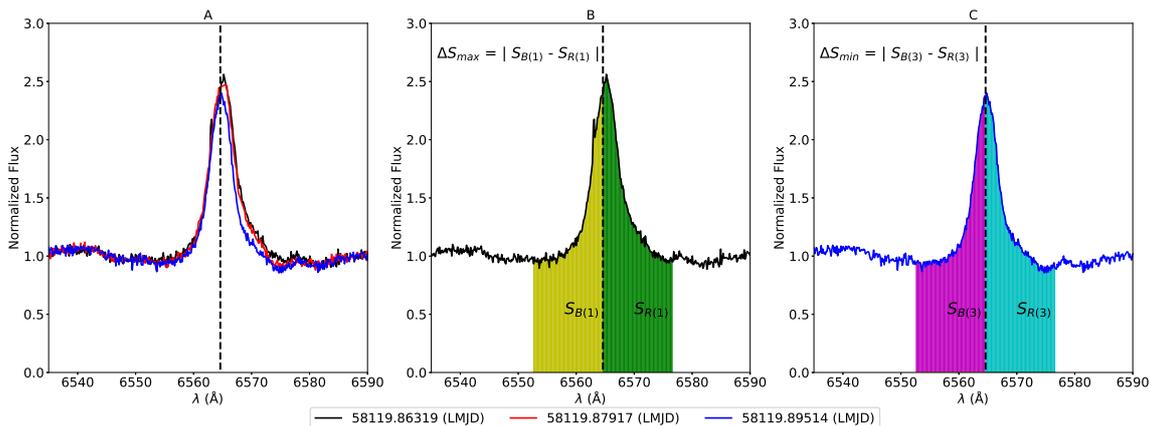}
      \caption{Schematic diagram of the partition integral comparing method (PICM) for stellar CME detection. Panel A shows the H$\alpha$ line profiles in a preprocessed file. Panels B and C show the maximum ($\Delta S_{max}$) and minimum values ($\Delta S_{min}$) of the $\Delta S_{n}$, respectively. The $\Delta S_{n}$ is the absolute value of the difference between the integrals on the blue and red halves of each H$\alpha$ profile. The black vertical dotted line represents the rest wavelength of the H$\alpha$ line in each panel.  }
         \label{CME_search}
   \end{figure*}       
   

\subsection{CME detection method}
  
   In this study, we searched for stellar CME signals by examing the asymmetry of the H$\alpha$ line. There are a number of techniques that can be used to analyze the asymmetries of line profiles \citep[e.g.][]{2011ApJ...738...18T, 2021PASJ...73...44M, 2021A&A...646A..34K}. In order to quickly and accurately select spectra with asymmetric H$\alpha$ line profiles from the large sample, we designed a partition integral comparing method (PICM) to analyze each file. Each file comprises at least three continuously observed spectra for a target.
    
   In the following we describe the principle of the PICM. Panel A of Fig. \ref{CME_search} shows the H$\alpha$ lines of the LAMOST-MRS spectra in a preprocessed file, which contains three continuously observed spectra. The exposure time of each spectrum is 1200 s. First, the blue region (6552.61 -- 6564.61 Å) and the red region (6564.61 -- 6576.61 Å) of the H$\alpha$ line in each spectrum were respectively integrated, as shown below.
     \begin{equation}
       S_{B(n)} = \int_{6552.61}^{6564.61} F_{nor}(\lambda)d\lambda, \qquad n=1,2,...,N
    \end{equation}
     \begin{equation}      
      err_{B(n)} = \sigma_{B(n)} \times 12, \qquad n=1,2,...,N
    \end{equation}   
    \begin{equation}     
    S_{R(n)} = \int_{6564.61}^{6576.61} F_{nor}(\lambda)d\lambda, \qquad n=1,2,...,N
    \end{equation}
    \begin{equation}  
    err_{R(n)} = \sigma_{R(n)} \times 12, \qquad n=1,2,...,N
    \end{equation}   
 In the above four equations, $S_{B(n)}$ and $S_{R(n)}$ are the integrals of the blue and red halves of the H$\alpha$ line for each spectrum in a file, respectively. $err_{B(n)}$ and $err_{R(n)}$ are the corresponding errors. And N is the number of spectra in the file. $F_{nor}$ is the normalized flux of the spectrum. $\sigma_{B(n)}$ and $\sigma_{R(n)}$ are the standard deviations of the normalized fluxes in the wavelength range of 6535 - 6545 Å and 6580 - 6590 Å, respectively. The integration limits for $S_{B(n)}$ and $S_{R(n)}$ are 12 Å, corresponding to a maximum Doppler velocity of 548 km s$^{-1}$. For most of the detected CME candidates, their H$\alpha$ profiles at least partially overlap with the integral regions. The disadvantage of this method is that very fast and very slow asymmetries may be missed when the CME-caused H$\alpha$ profile is narrow.
 
     We then calculated the absolute value of the difference between the integrals for each spectrum using the following formula.
    \begin{equation}      
    \Delta S_{n} = \left | S_{B(n)} - S_{R(n)} \right |, \qquad n=1,2,...,N
    \end{equation}
    \begin{equation}  
   err_{S_{n}} = \sqrt{(err_{B(n)})^{2} + (err_{R(n)})^{2}}, \qquad n=1,2,...,N
     \end{equation}   
When $\Delta S_{n}$ is greater than its corresponding error $err_{S_{n}}$, it indicates that the H$\alpha$ line in the spectrum is asymmetric. As illustrated in panels B and C of Fig. \ref{CME_search}, the maximum value of $\Delta S_{n}$ is represented by $\Delta S_{max}$ and the minimum value is represented by $\Delta S_{min}$. In order to ensure that the asymmetric H$\alpha$ profiles originate from the process of stellar eruption, rather than from the quiet stellar prominence, we further imposed the following requirement: 
    \begin{equation}       
     \left | \Delta S_{max} - \Delta S_{min} \right | > \sqrt{(err_{S_{max}})^{2} + (err_{S_{min}})^{2}}
     \end{equation}        
If the H$\alpha$ line profiles in a file satisfy the above conditions, the PICM will treat the file as a stellar CME candidate. Finally, we visually inspected the automatic search result to remove false stellar CME candidates caused by residual cosmic rays. In the meantime, we checked the SIMBAD database \citep{2000A&AS..143....9W} to ensure that each CME candidate comes from a late-type main-sequence single star, excluding the influence of eclipsing binaries, T Tauri stars, and RR Lyr stars.      
   
   After these processes, we detected three stellar CME candidates, and their related parameters are listed in Table \ref{table:2}. In Table \ref{table:2}, the first column is "LAMOST obsid", representing the unique identification of each file in which a CME candidate was identified. The second column is "LAMOST designation", which is the LAMOST name of the host star for each CME candidate. The third column is the spectral type of the host stars, which is classified by the LAMOST 1D pipeline by matching the observed low-resolution spectrum with the templates \citep{2015RAA....15.1095L}. The uncertainty of the LAMOST spectral type is within two subclasses. The fourth and fifth columns are the stellar effective temperature and surface gravity given by the LAMOST-MRS. The sixth and seventh columns are stellar parameters from the Two Micron All Sky Survey (2MASS), including the 2MASS designation and the J-band magnitude \citep{2003yCat.2246....0C}. The last column is the stellar radius calculated from the relationship between the radius and spectral type for main-sequence stars \citep{2000asqu.book.....C}. The locations of the host stars of these three CME candidates on the Hertzsprung-Russell diagram are shown as the red solid circles in Fig. \ref{HR}.

%
%

\begin{table*}
\caption{Numbers of stars in our search sample.}             
\label{table:1}      
\centering 
\begin{tabular}{c c c c c}     
\hline\hline       
                     
Data name & G-type stars (spectra) & K-type stars (spectra) & M-type stars (spectra) & All stars (spectra)      \\ 
\hline    
 LAMOST DR8  &  138,928 (863,794) & 30,201 (181,858) & 1393 (7103) & 170,522 (1,052,755) \\      
LAMOST DR9 v0 &  44,889 (265,540) & 10,355 (59,075) & 428 (2038) & 55,672 (326,653) \\      
 CME candidates &  0 (0) & 0 (0) & 3 (11) & 3 (11)   \\      
\hline                  
\end{tabular}
\end{table*}

\begin{table*}
\caption{Observation information and stellar parameters for the three CME candidates.}             
\label{table:2}      
\centering
\begin{tabular}{cccccccc}     
\hline\hline       
                     
LAMOST obsid& LAMOST designation & Sp\_type & $T_{\mathrm{eff}}$(K) & $\log g$(dex) & 2MASS designation & J(mag) & Radius($R_{\sun}$)       \\ 
\hline    
876604049 &  J035012.86+242106.5 & M1 & 3708.7 & 4.7 &  J03501290+2421067 & 11.454 & 0.55 \\         
635003103 &  J121933.15+015426.7 & M4 & 3436.3 & 4.6 & J12193316+0154268 &  10.543 & 0.35 \\  
624510064 &  J041827.35+145813.6 & M2 & 3540.1 & 4.8 & J04182735+1458137 & 10.444 & 0.50 \\     
\hline                  
\end{tabular}
\end{table*}

\section{Results and discussion}

\subsection{Stellar CME candidate 1: LAMOST obsid 876604049}

\subsubsection{Characteristics of the chromospheric lines in CME candidate 1}

 
  \begin{figure*}
   \centering
   \includegraphics[width=\hsize]{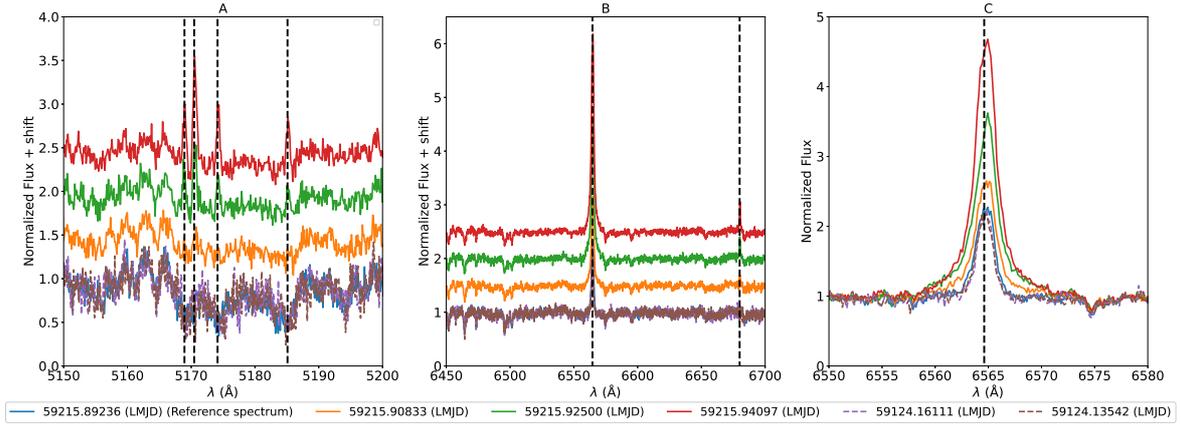}
      \caption{The preprocessed spectra of CME candidate 1 (four solid lines) and the reference spectra of the host star (two dashed lines). Each spectrum is marked with the LAMOST Local Modified Julian Day (LMJD) at the time of exposure.  Panel A is part of the blue arm (5150 -- 5200 Å), which contains the Mg\,{\sc{i}} triplet lines (5168.94 Å, 5174.13 Å, 5185.10 Å) and the Fe\,{\sc{ii}} 5170.47Å line. The rest wavelengths of these chromospheric lines are marked with the black vertical dotted lines. Panel B is part of the red arm (6450 -- 6700 Å), which contains the H$\alpha$ (6564.61 Å) line and the He\,{\sc{i}} 6680Å line. The vacuum wavelengths of these lines are taken from \cite{2018Galax...6...63V}. In panels A and B, the spectra of CME candidate 1 observed at different times are shifted on the vertical axis for a better illustration. In order to clearly show the changes of the H$\alpha$ line profiles, the spectra near the H$\alpha$ line are plotted in panel C. }
         \label{CME_1_spectra}
   \end{figure*}
 
  
   \begin{figure*}
   \centering
   \includegraphics[width=\hsize]{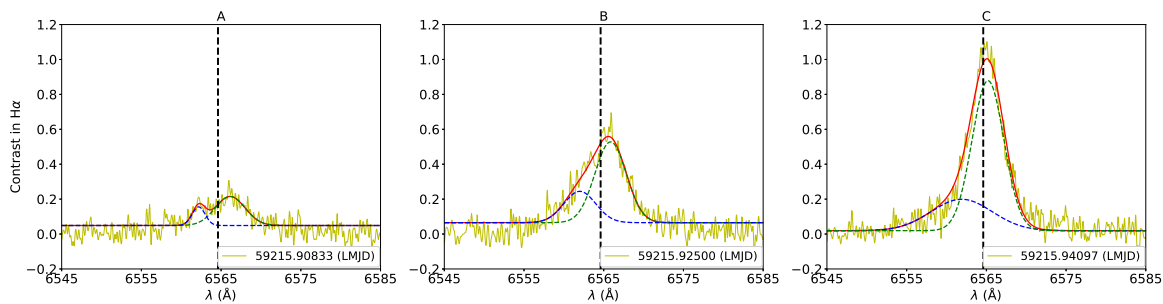}
      \caption{Double Gaussian fitting results for the H$\alpha$ contrast profiles of the three active spectra in CME candidate 1. The yellow solid lines are the H$\alpha$ contrast profiles, the red solid lines are the double Gaussian fitting results. The blue and green dashed lines represent the two Gaussian components. The black vertical dotted line indicates the rest wavelength of the H$\alpha$ line in each panel.    }
         \label{CME_1_Hafitting}
   \end{figure*}
 
   
   \begin{figure*}
   \centering
   \includegraphics[width=\hsize]{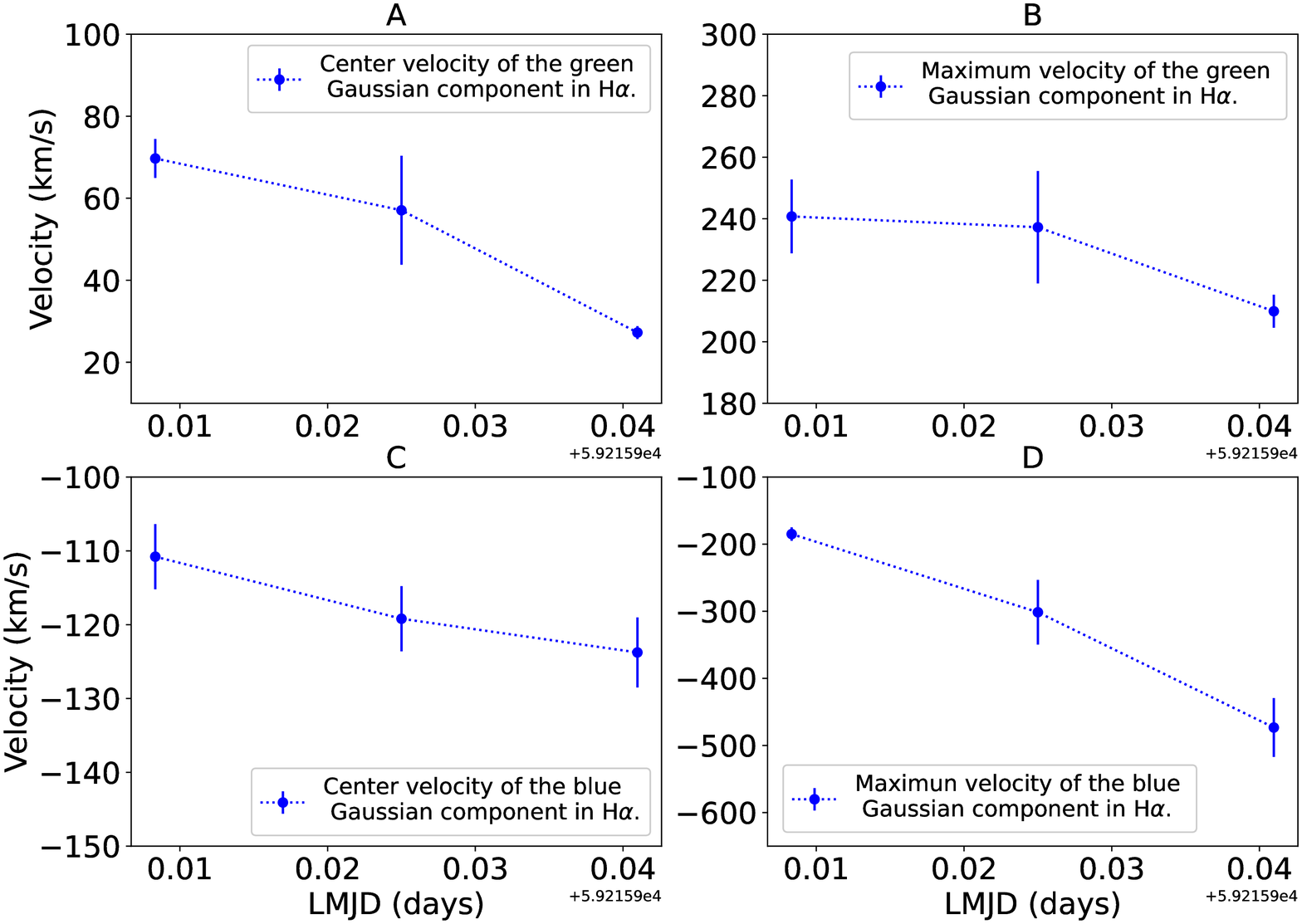}
      \caption{The center and maximum velocities of the two components in the H$\alpha$ contrast profiles of the three active spectra in CME candidate 1. Panel A presents the center velocities of the components shown by the green dashed lines in Fig. \ref{CME_1_Hafitting} and panel B shows the corresponding maximum velocities. Panel C presents the center velocities of the components shown by the blue dashed lines in Fig. \ref{CME_1_Hafitting} and panel D shows the corresponding maximum velocities. The error bars represent the double Gaussian fitting errors. }
         \label{CME_1_Havelocity}
   \end{figure*}

 \begin{figure}
   \centering
     \includegraphics[width=\hsize]{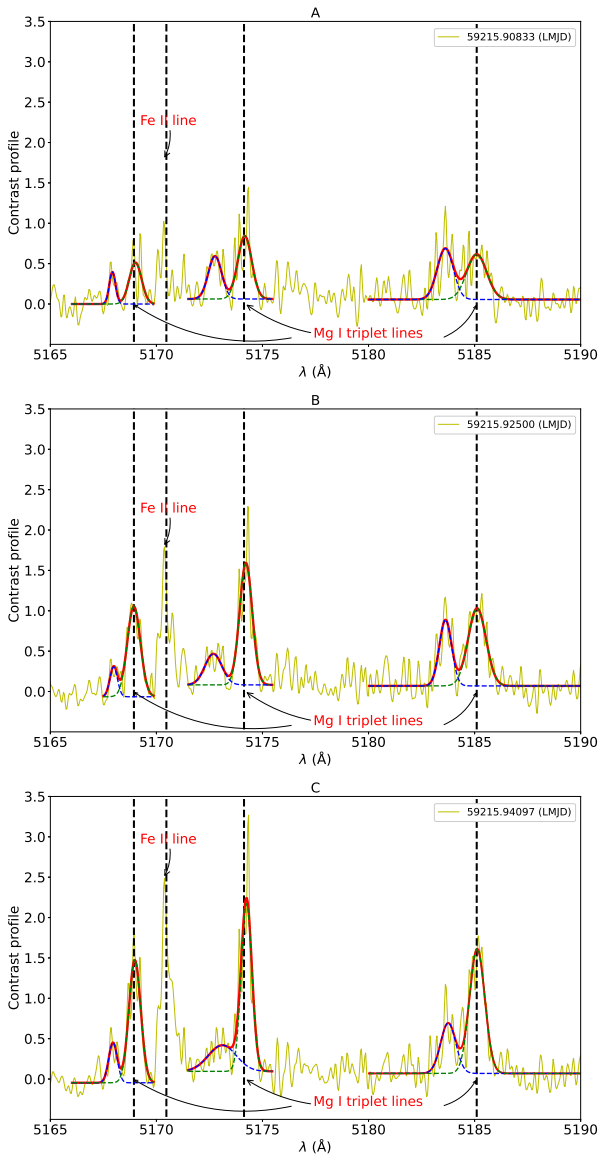}
      \caption{Double Gaussian fitting results for the contrast profiles of the three active spectra in CME candidate 1. The black vertical dotted lines indicate the rest wavelengths of the Mg\,{\sc{i}} triplet and Fe\,{\sc{ii}} 5170.47Å lines.    }
         \label{CME_1_MgIfitting}
  \end{figure}
  
   \begin{figure*}
   \centering
     \includegraphics[width=\hsize]{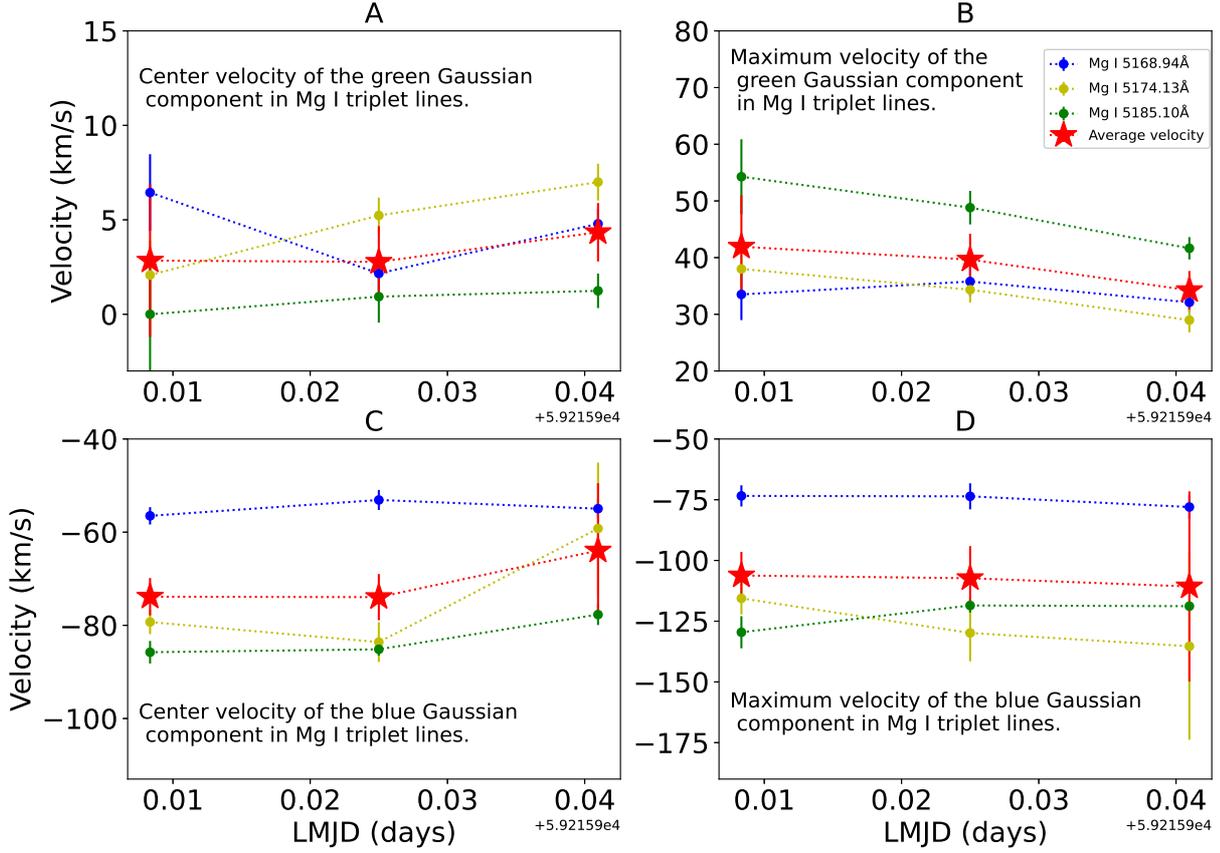}  
      \caption{The center and maximum velocities of the two components in the contrast profiles of the Mg\,{\sc{i}} triplet lines in CME candidate 1. Panel A presents the center velocities of the components shown by the green dashed lines in Fig. \ref{CME_1_MgIfitting}, and panel B shows the corresponding maximum velocities. Panel C presents the center velocities of the components shown by the blue dashed lines in Fig. \ref{CME_1_MgIfitting} and panel D shows the corresponding maximum velocities. The blue, yellow and green solid circles represent velocities of the Mg\,{\sc{i}} triplet lines, respectively. And the red stars represent the corresponding average velocities.     }
         \label{CME_1_MgIvelocity}
   \end{figure*}

   \begin{figure*}
   \centering
     \includegraphics[width=\hsize]{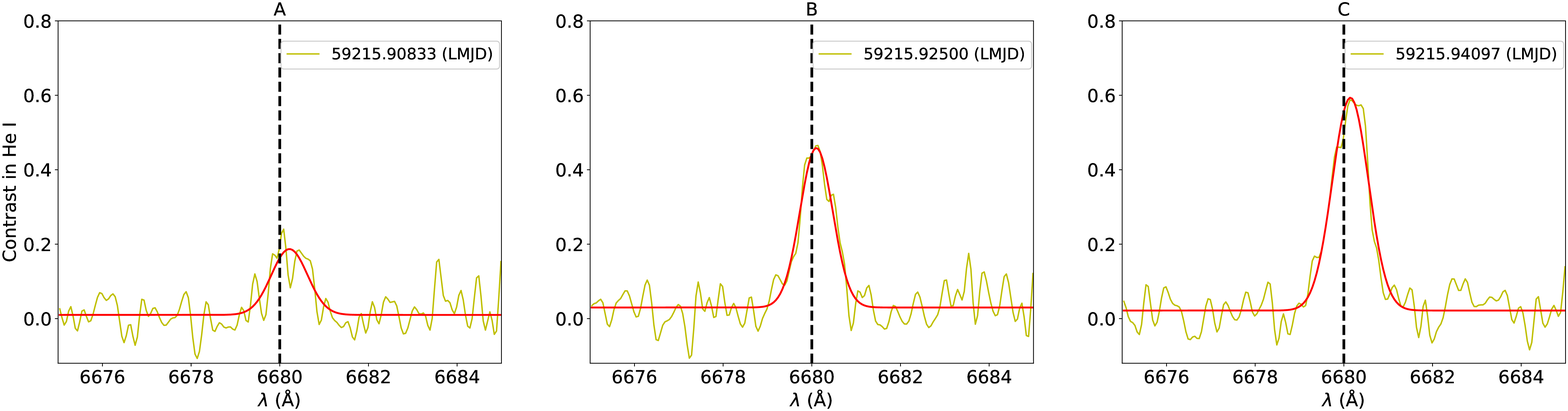}  
      \caption{Single Gaussian fitting results for the He\,{\sc{i}} 6680Å contrast profiles of the three active spectra in CME candidate 1. The red solid lines are the fitting results, and the black vertical dotted lines indicate the rest wavelength of the He\,{\sc{i}} 6680Å line.   }
         \label{CME_1_HeIIfitting}
   \end{figure*}
 
 
    The host star of CME candidate 1 (LAMOST obsid 876604049) is a main-sequence star (LAMOST J035012.86+242106.5). This host star is also named V371 Tau in the Pleiades. Several authors have confirmed that V371 Tau is a dwarf with a spectral type of M3Ve \citep{1961ApJ...133.1085P, 1990Afz....33....5M, 1991AJ....101.1361P, 1993Afz....36..501P}. This star is also a flare star \citep{1970BOTT....5...23H, 1975Ap.....11..565C, 1982BITon...3....3H, 1990Afz....33....5M, 1993Afz....36..501P, 1995MNRAS.274..869H, 2001Ap.....44..106A}. There are four consecutively observed LAMOST-MRS spectra for CME candidate 1. Fig. \ref{CME_1_spectra} shows the preprocessed spectra of CME candidate 1 and two reference spectra of its host star. Following \cite{2021A&A...646A..34K}, when the normalized spectra of the host star observed at different times are almost the same, these spectra are regarded as reference spectra. The first spectrum of CME candidate 1 is almost the same as the two reference spectra of the host star, indicating that this spectrum was also taken in the relatively quiet period. So we superimposed these three spectra and plotted them in Fig. \ref{CME_1_spectra}. It can be seen from Fig. \ref{CME_1_spectra} that the intensities of the chromospheric lines (Mg\,{\sc{i}} triplet lines, Fe\,{\sc{ii}} 5170.47Å, H$\alpha$ and He\,{\sc{i}} 6680Å) in the second, third and fourth spectra (orange, green and red solid lines) of CME candidate 1 gradually increase, indicating the impulsive phase of a stellar flare. In order to analyze the asymmetries of the chromospheric lines in detail, we defined the following contrast profile.
    \begin{equation}
       I_{c}(\lambda) = \frac{F_{nor, active}(\lambda ) - F_{nor, ref}(\lambda )}{F_{nor, ref}(\lambda )},
    \end{equation}   
where $F_{nor, active}(\lambda )$ is the normalized flux of an active spectrum in the CME candidate, and $F_{nor, ref}(\lambda )$ is the normalized flux of the reference spectrum. The active spectrum and the reference spectrum were interpolated before this step. This operation allows these spectra to be compared at the same wavelengths. The contrast profile is often used to analyze asymmetric spectral line profiles observed during solar magnetic activity \citep{2014ApJ...792...13H}. For CME candidate 1, the first spectrum is a reference spectrum and has a S/N of 23 for the red arm and 8 for the blue arm. Therefore, this spectrum was selected as the reference spectrum and marked as "Reference spectrum" in Fig. \ref{CME_1_spectra}. The three H$\alpha$ contrast profiles of the active spectra are shown by the yellow solid lines in Fig. \ref{CME_1_Hafitting}.

   In this study, we performed double Gaussian fitting for the contrast profiles of the asymmetric chromospheric lines, the fitting function can be expressed as the following,
   \begin{equation}
      I_{c}(\lambda) = A_{0} + A_{1} e^{-\frac{(\lambda - \lambda _{1})^{2}}{2 \sigma _{1}^{2}}} + A_{2} e^{-\frac{(\lambda - \lambda _{2})^{2}}{2 \sigma _{2}^{2}}}
   \end{equation}
In addition, we also defined the center and maximum velocities of the two Gaussian components, 
    \begin{equation}
       \upsilon_{1,center} = \frac{\lambda_{1} - \lambda_{0}}{\lambda_{0}}c\,,
    \end{equation}    
    \begin{equation}
       \upsilon_{1,max} = \frac{\lambda_{1} - \lambda_{0} \pm 2 \sigma _{1}}{\lambda_{0}}c\,,
    \end{equation}    
    \begin{equation}
       \upsilon_{2,center} = \frac{\lambda_{2} - \lambda_{0}}{\lambda_{0}}c\,,
    \end{equation}    
    \begin{equation}
       \upsilon_{2,max} = \frac{\lambda_{2} - \lambda_{0} \pm 2 \sigma _{2}}{\lambda_{0}}c\,,
    \end{equation}    
where $\lambda_{0}$ is the centroid wavelength of a chromospheric line, c is the speed of light. The sign "$\pm$" in equations (11) and (13) is determined by the results of equations (10) and (12), respectively. When the centroid velocity is negative, it is taken as "-", otherwise it is taken as "+". When the speed is negative, it means a blue shift. And when it is positive, it indicates a red shift.    

     The double Gaussian fitting results of the H$\alpha$ contrast profiles of the three active spectra in CME candidate 1 are presented in Fig. \ref{CME_1_Hafitting}. The fitting results reveal that there are two components in the H$\alpha$ contrast profiles, one of which is the blueshifted component, as shown by the blue dashed lines in Fig. \ref{CME_1_Hafitting}, and the other is the redshifted component shown by the green dashed lines. The amplitude and the full width at half maximum (FWHM) of these two components are gradually increasing. In addition, it can be seen from Fig. \ref{CME_1_Havelocity} that the central and maximum velocities of the blueshifted component both show a gradually increasing trend, while the central and maximum velocities of the redshifted component show a gradually decreasing trend.      
     
      For CME candidate 1, the S/N values of the red arms of the three active spectra are 25, 24, 27, respectively. The values of the blue arms are 10, 9, 11, respectively. Although the S/N of the blue arm is not as high as that of the red arm, it can be seen from the contrast profiles shown in Fig. \ref{CME_1_MgIfitting} that the emission of the Mg\,{\sc{i}} triplet lines gradually increases, and their blue wings also show obvious enhancements. Therefore, we also performed double Gaussian fitting for the contrast profiles of the Mg\,{\sc{i}} triplet lines, and present the results in Fig. \ref{CME_1_MgIfitting}. The Fe\,{\sc{ii}} 5170.47Å line emission can also be seen from Fig. \ref{CME_1_MgIfitting} and its intensity gradually increases. Fig. \ref{CME_1_MgIvelocity} shows the center and maximum velocities of the two components of the Mg\,{\sc{i}} triplet lines. From panel A we can see that the components marked by the green dashed lines in Fig. \ref{CME_1_MgIfitting} show a small red shift, with an average value of the central velocities around 5 km s$^{-1}$. The panel B shows that their maximum velocities are decreasing. The components marked by the blue dashed lines in Fig. \ref{CME_1_MgIfitting} reveal a blue shift. The average value of the center velocities of these blueshifted components is about --70 km s$^{-1}$ (panel C), and the average value of the maximum velocities is about --110 km s$^{-1}$ (panel D). The single Gaussian fitting results for the He\,{\sc{i}} 6680Å contrast profiles are presented in Fig. \ref{CME_1_HeIIfitting}. We can see that the Doppler shift is small and that the most obvious feature is the gradually increasing intensity.

\subsubsection{Asymmetries of the chromospheric lines in CME candidate 1}

   The above analysis results indicate that CME candidate 1 is likely detected in the impulsive phase of the associated stellar flare. More importantly, the Mg\,{\sc{i}} triplet lines and the H$\alpha$ line show blue-wing enhancements at the same time. This is the first time that a blue-wing enhancement of Mg\,{\sc{i}} triplet lines has been detected during a stellar flare. By checking the atomic line database \citep{2018Galax...6...63V}, no flare-related lines were found to appear in the blue wings of the Mg\,{\sc{i}} triplet lines simultaneously. The S/N of the H$\alpha$ is twice that of the Mg\,{\sc{i}} triplet lines, and the center velocity and maximum velocity of the blue-wing enhancement in H$\alpha$ clearly show a gradually increasing trend. Therefore, the blue-wing enhancement could be interpreted as the line-of-sight projection of an outward moving stellar prominence (as part of a CME). The following is the basis of this interpretation. 
   
   First, \cite{2017ApJ...842L...7D} observed the spectra of a solar CME associated with a prominence eruption during a total solar eclipse. Their spectra show an obvious emission in the Mg\,{\sc{i}} triplet, Fe\,{\sc{ii}} and He\,{\sc{i}} lines. The redshifts of these lines correspond to velocities ranging from under 100 to over 1500 km s$^{-1}$. Blueshifts are rarely found in their lines. These observational results were interpreted as being caused by a solar CME moving away from the observer. The blue-wing enhancement of the spectral lines in our data may be caused by a CME propagating towards the observer. It should be noted that the blue-wing enhancement of the Mg\,{\sc{i}} triplet lines is noisy and seems to be always separated from the line core. The corresponding velocities are small. Further observations are required to examine whether such features are CME-related. 
   
   Second, is it possible that these blue-wing enhancements are caused by the chromospheric evaporation? Solar flare observations show that signatures of chromospheric evaporation are often found in high-temperature lines, such as emission lines from the Fe\,{\sc{xii}}-{\sc{xxiv}} ions, while low-temperature chromospheric lines often do not reveal obvious blue-wing enhancements or blue shifts \citep[e.g.,][]{2014ApJ...797L..14T, 2015ApJ...811..139T, 2011ApJ...727...98L, 2015ApJ...811....7L, 2015ApJ...813...59L}. In a solar flare observation, \cite{2018PASJ...70..100T} found that the Mg\,{\sc{ii}} h line shows a blue-wing enhancement in the impulsive phase of the flare. After using the cloud model to analyze the evolution of the Mg\,{\sc{ii}} h line profile during the flare, they suggested that the blue-wing enhancement of the Mg\,{\sc{ii}} h line may be caused by a scenario in which an upflow of cool plasma is lifted up by expanding hot plasma owing to the deep penetration of non-thermal electrons into the chromosphere. However, the maximum speed corresponding to the blue-wing enhancement of the Mg\,{\sc{ii}} h is only 14 km s$^{-1}$, which is much smaller than that in our case. We noticed that the velocities of the blue-wing enhancement of the Mg\,{\sc{i}} triplet lines and the H$\alpha$ line in CME candidate 1 are different, which might be caused by the different S/N of these lines or the larger uncertainty of the H$\alpha$ fitting. There is no obvious blue-wing enhancement in the Fe\,{\sc{ii}} 5170.47 \AA~and He\,{\sc{i}} 6680 \AA~lines, which may be caused by the low S/N in these lines. However, the enhanced emission in both of these two spectral lines indicates that the host star is indeed in the process of an eruption.         

   The redshifted component in the H$\alpha$ contrast profiles may be caused by chromospheric condensation. A solar flare observation reveals that the red shift of the H$\alpha$ line caused by chromospheric condensation reaches 46 km s$^{-1}$ before gradually decreasing \citep{2018PASJ...70..100T}. The decreasing velocity of the redshifted component of H$\alpha$ in CME candidate 1 is similar to the result of \cite{2018PASJ...70..100T}, and the center velocity is close to the velocity of chromospheric condensation in the solar observation. The maximum velocity of the redshifted component in the Mg\,{\sc{i}} triplet lines also shows a gradually decreasing trend. The central velocity shows a red shift that is smaller than that of H$\alpha$, which might be due to the different formation heights as well as S/N values of the H$\alpha$ and Mg\,{\sc{i}} triplet lines.
   
   If the blue-wing enhancements of the Mg\,{\sc{i}} triplet and H$\alpha$ lines in CME candidate 1 are caused by a CME, we can select the strongest blue wing enhancement of the H$\alpha$ line to estimate the minimum mass of the CME through the following formula \citep{1990A&A...238..249H, 2021A&A...646A..34K},
    \begin{equation}
       M_{CME} \geqslant \frac{4 \pi R^{2} F_{emission} \frac{N_{total}}{N_{j}} m_{H} \eta_{OD} }{h \nu_{j-i}A_{j-i}}\,,
    \end{equation}        
where $R$ is the stellar radius, $F_{emission}$ is the integrated flux of the H$\alpha$ blue wing enhancement, and other parameters are specifically referred to \cite{2021A&A...646A..34K}. Because the LAMOST-MRS only gives the relative flux of the spectrum, we calculated the $F_{emission}$ using the following method. We first integrated the enhancement area caused by the CME in the H$\alpha$ contrast profile, denoted as $Q_{CME}$. This represents the ratio of the H$\alpha$ enhancement caused by the CME to the quiet H$\alpha$ flux, which can be calculated using the following equation,
  \begin{equation}
       Q_{CME} = A_{c} \sigma _{c} \sqrt{2\pi }\,,
    \end{equation}    
where $A_{c}$ and $\sigma _{c}$ are the amplitude and standard deviation of the blueshifted Gaussian component of the H$\alpha$ contrast profile (blue dashed line in Fig. \ref{CME_1_Hafitting}C), respectively. The quiet H$\alpha$ flux ($F_{H\alpha , quiet}$) was calculated through the following empirical relationship,
  \begin{equation}
F_{H\alpha , quiet} \approx F_{H\alpha , con}=\chi \sigma T_{eff}^{4},
  \end{equation}    
where $\chi$ is the ratio of the surface continuum flux (adjacent to the H$\alpha$ line, $F_{H\alpha , con}$) to the stellar surface bolometric flux, which is given by \cite{2018MNRAS.476..908F} when calculating the flux of H$\alpha$ line in the LAMOST spectra. $\sigma$ is the Stefan-Boltzmann constant. Subsequently, we calculated the $F_{emission}$ through the following equation,
    \begin{equation}
       F_{emission} = Q_{CME} F_{H\alpha,quiet}\,.
    \end{equation}       
Finally, the minimum mass of CME candidate 1 was estimated to be $4.12 \times 10^{18}$ g, which is within the mass range of CME candidates for M dwarfs reported by \cite{2019A&A...623A..49V} and \cite{2021PASJ...73...44M}.

\subsection{Stellar CME candidate 2: LAMOST obsid 635003103}


 \begin{figure*}
   \centering
   \includegraphics[width=\hsize]{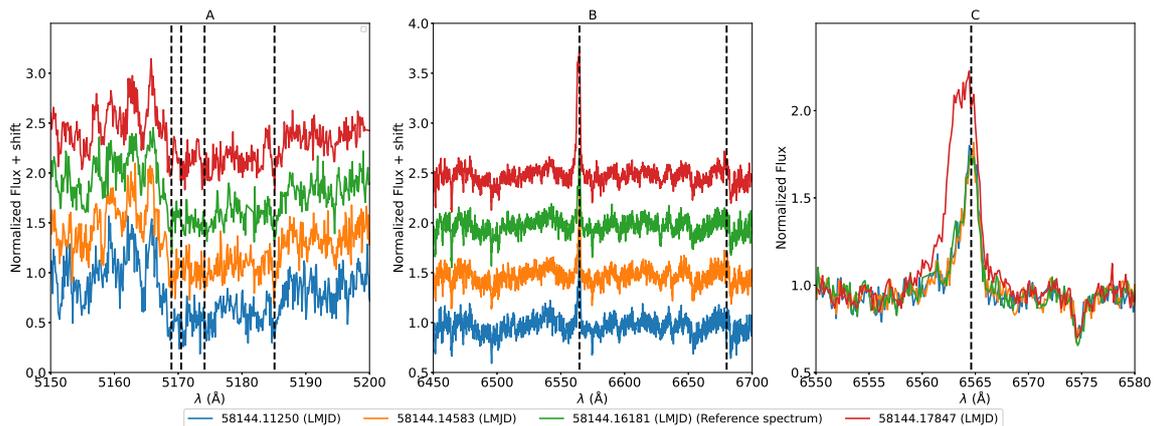}
      \caption{Four preprocessed LAMOST-MRS spectra of CME candidate 2. The first three spectra (blue, orange and green solid lines) are almost the same and they are regarded as reference spectra of the host star. The three panels are similar to those in Fig. \ref{CME_1_spectra}.     }
         \label{CME_3_spectra}
   \end{figure*}

 
  \begin{figure*}
   \centering
   \includegraphics[width=\hsize]{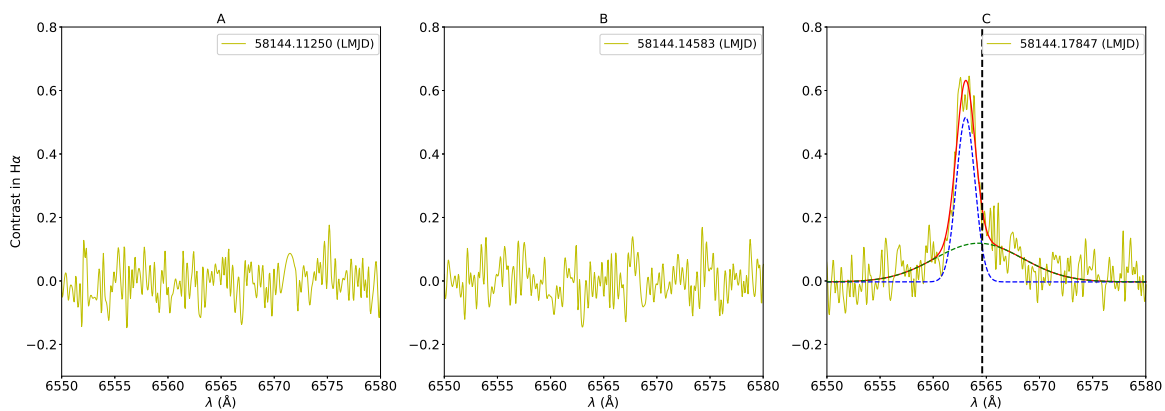}
      \caption{ H$\alpha$ contrast profiles in CME candidate 2. Panels A and B show the H$\alpha$ contrast profiles of the first two spectra. Panel C shows the double Gaussian fitting result for the H$\alpha$ contrast profile of the active spectrum.     }
         \label{CME_3_Hafitting}
   \end{figure*}
   

   The host star of CME candidate 2 (LAMOST obsid 635003103) is an M4-type main-sequence star (LAMOST J121933.15+015426.7). There are four consecutively observed LAMOST-MRS spectra for CME candidate 2. Fig. \ref{CME_3_spectra} shows the four preprocessed spectra. Although the H$\alpha$ line profiles in the first three spectra appear to show a slight asymmetry, these three spectra are almost the same. We thus regarded these three spectra as reference spectra of the host star. The S/N values of the blue and red arms of the third spectrum are 7 and 15, respectively. They are both higher than those of the other two spectra. Thus, we chose the third spectrum as the reference spectrum and performed double Gaussian fitting for the H$\alpha$ contrast profile of the active spectrum. The result is shown in Fig. \ref{CME_3_Hafitting}. Panels A and B of Fig. \ref{CME_3_Hafitting} show that the H$\alpha$ contrast profiles of the first two spectra reveal no obvious asymmetries. The double Gaussian fitting result in panel C shows that the H$\alpha$ contrast profile exhibits a blueshifted component and a wide emission component centered around the rest wavelength of the H$\alpha$ line. The center velocity of the blueshifted component is --71 km s$^{-1}$ and the maximum velocity is --151 km s$^{-1}$. The blueshifted component in the H$\alpha$ contrast profile may be caused by the projection of a CME eruption in the line of sight. Although its center velocity is low, it is in the impulsive phase of the eruption and may continue to accelerate in the later stage, thereby perhaps escaping from the host star. The minimum mass of this possible CME was estimated to be $8.84 \times 10^{17}$ g. The large line width of the broad emission component may result from the Stark broadening associated with the enhanced pressure by electrons \citep{1972SoPh...24..154S, 2019ApJ...879...19Z, 2020A&A...637A..13M, 2022ApJ...928..180W}.

\subsection{Stellar CME candidate 3: LAMOST obsid 624510064}

\subsubsection{Characteristics of the chromospheric lines in CME candidate 3}

   \begin{figure*}
   \centering
   \includegraphics[width=\hsize]{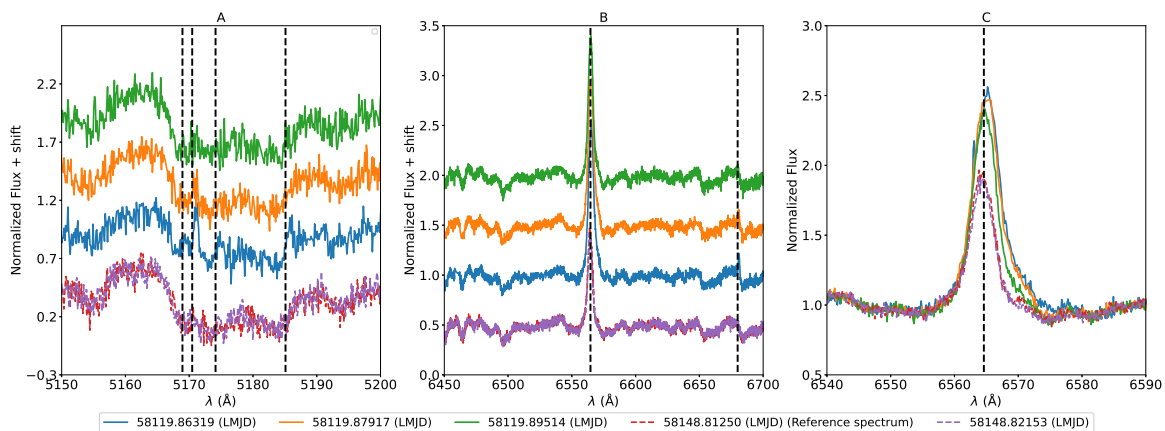}
      \caption{Three preprocessed spectra of CME candidate 3 (three solid lines) and two reference spectra of the host star (two dashed lines).  The three panels are similar to those in Fig. \ref{CME_1_spectra}.   }
         \label{CME_4_spectra}
   \end{figure*}
   \begin{figure*}
   \centering
   \includegraphics[width=\hsize]{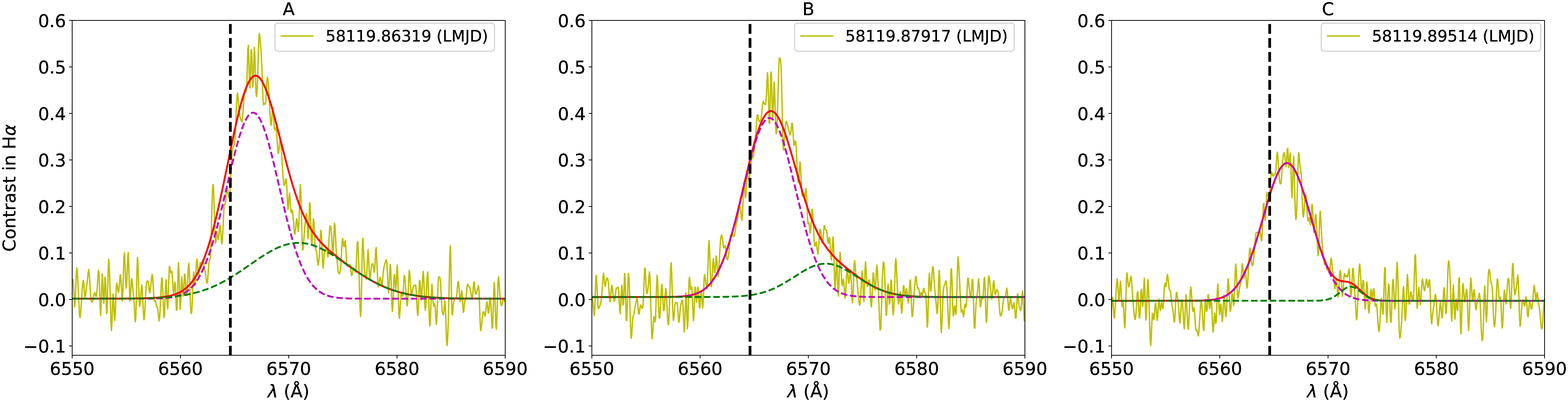}
     \includegraphics[width=\hsize]{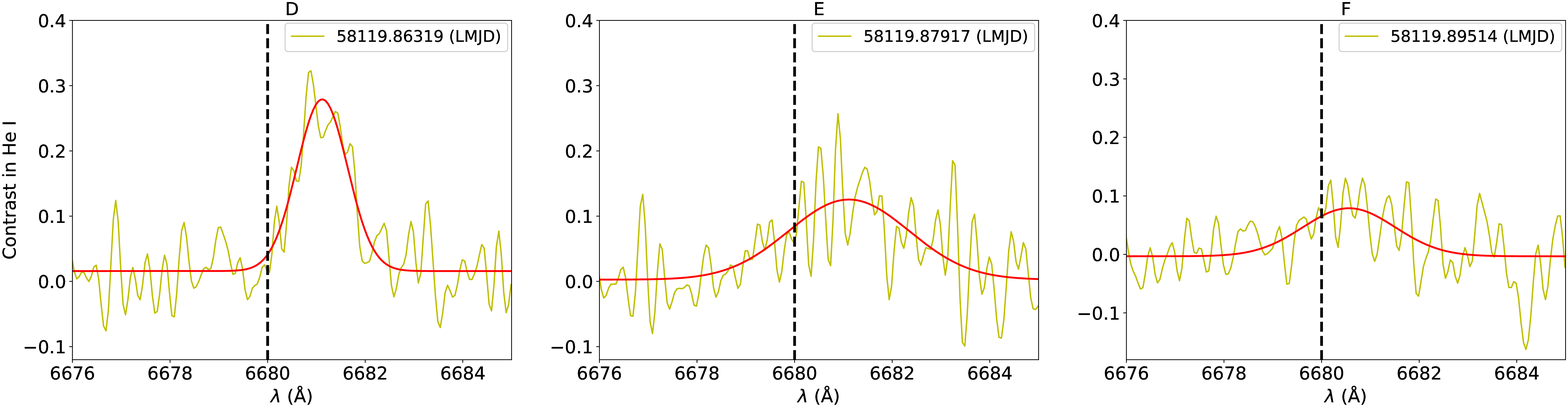}  
     \includegraphics[width=10cm]{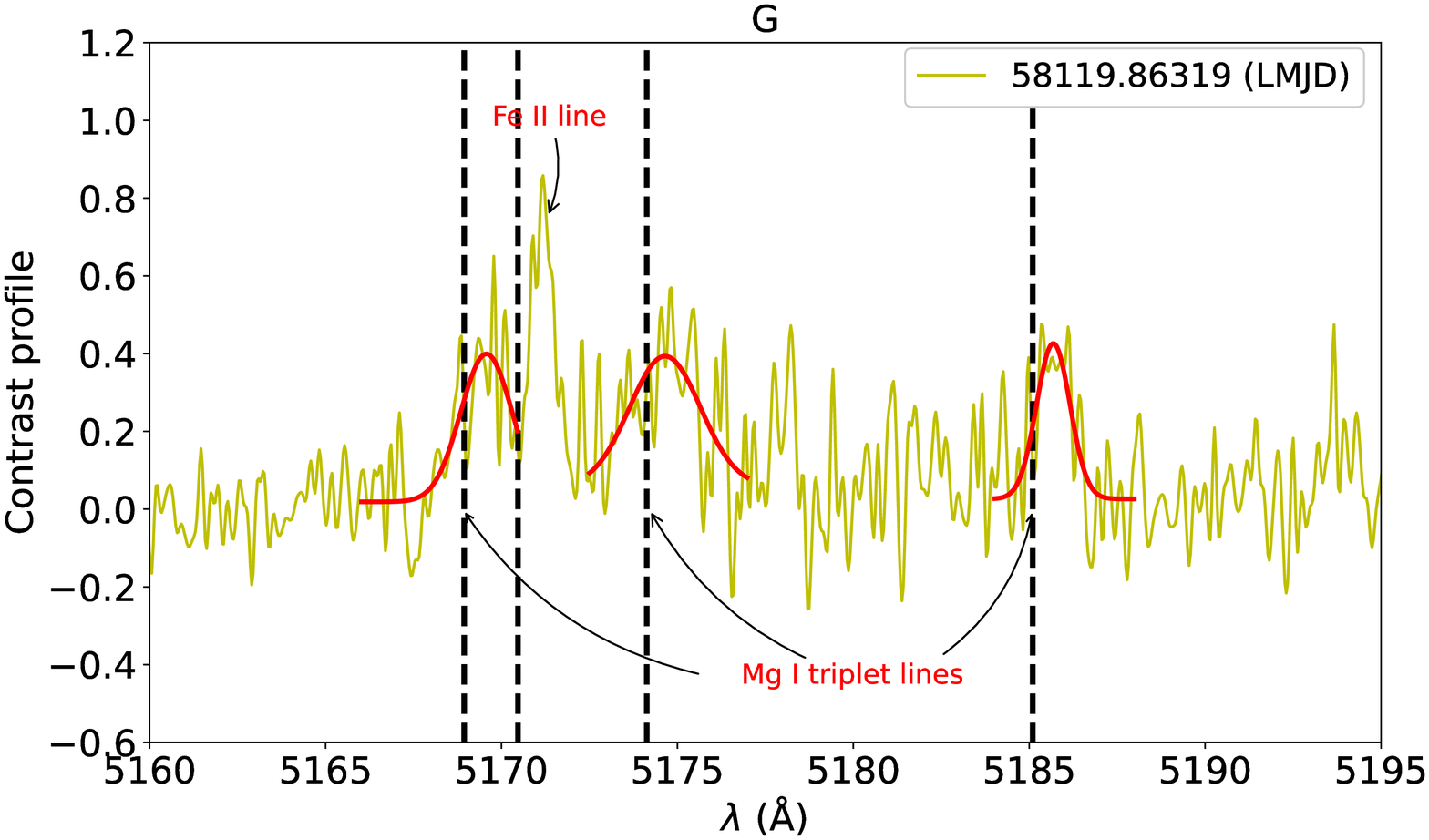}
      \caption{Gaussian fitting results for the contrast profiles of the H$\alpha$, He\,{\sc{i}} 6680Å and Mg\,{\sc{i}} triplet lines in the active spectra of CME candidate 3. Panels A, B and C are the results for H$\alpha$. The red solid lines are the double Gaussian fitting results, the pink and green dashed lines represent the two Gaussian components. Panels D, E and F present the single Gaussian fitting results for the He\,{\sc{i}} 6680Å line, and the red solid lines are the fitting results. Panel G presents the single Gaussian fitting results for the Mg\,{\sc{i}} triplet lines in the first active spectrum (blue solid line in Fig. \ref{CME_4_spectra}).     }
         \label{CME_4_linefitting}
   \end{figure*}

 
   \begin{figure*}
   \centering
   \includegraphics[width=\hsize]{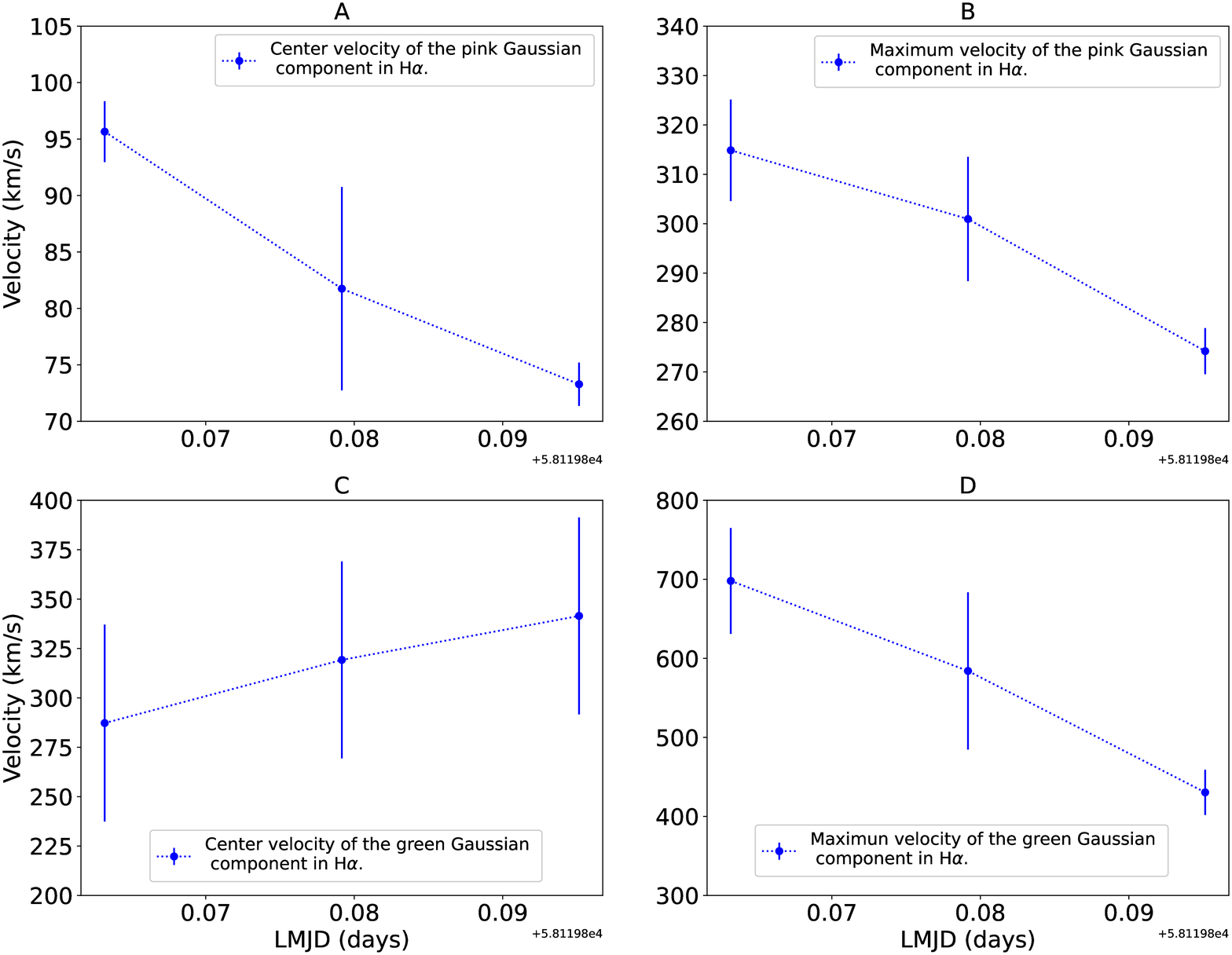}
      \caption{Temporal evolution of the center and maximum velocities of the two components in the H$\alpha$ contrast profiles of the three active spectra for CME candidate 3. }
         \label{CME_4_Havelocity}
   \end{figure*}
 
    \begin{figure*}
   \centering
   \includegraphics[width=\hsize]{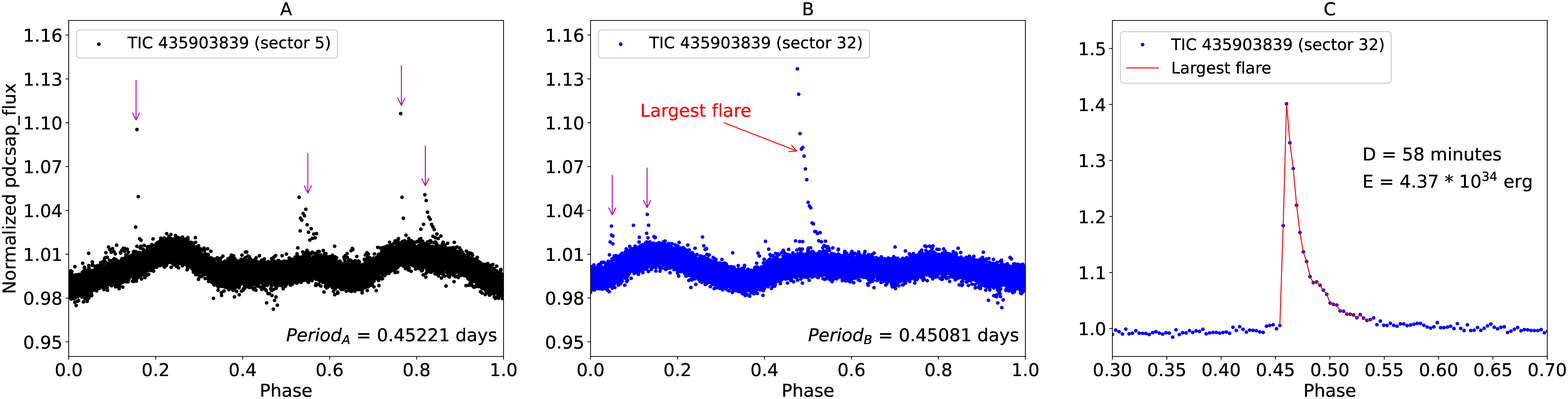}
      \caption{TESS photometric data for the host star of CME Candidate 3. Panel A shows the photometric data of TESS sector 5. The observation time is from November 15 to December 11, 2018. Several superflares are indicated by the pink arrows. Panel B shows the photometric data of TESS sector 32. Its observation time is from November 19 to December 17, 2020. The largest superflare in the TESS observation is marked by the red arrow. Panel C shows the zoomed-in version of the light curve of the largest superflare. "D" and "E" represent the flare duration and energy, respectively.         }
         \label{CME_4_TESS}
   \end{figure*}
 
 
   The host star of CME candidate 3 (LAMOST obsid 624510064) is an M2-type main-sequence star (LAMOST J041827.35+145813.6). There are three consecutively observed LAMOST-MRS spectra for CME candidate 3. Fig. \ref{CME_4_spectra} shows the three preprocessed spectra of CME candidate 3 and two reference spectra of the host star. We can see that the H$\alpha$ emission in the active spectra gradually weakens and the red wing of H$\alpha$ shows an obvious enhancement, possibly indicating the decay phase of a stellar flare. In addition, the Mg\,{\sc{i}} triplet, Fe\,{\sc{ii}} 5170.47Å  and He\,{\sc{i}} 6680Å lines also exhibit emission. The double Gaussian fitting results for the H$\alpha$ contrast profiles of the three active spectra in CME candidate 3 are shown in panels A, B and C of Fig. \ref{CME_4_linefitting}. There are two redshifted components in the H$\alpha$ contrast profiles, and both of their intensities gradually decrease.  Fig. \ref{CME_4_Havelocity} shows that the center and maximum velocities of the low-velocity component gradually decrease. The center velocity of the high-velocity component shows a slightly increasing trend, but we realize that the fitting error is large. The maximum velocity gradually decreases and the largest velocity reaches 698 km s$^{-1}$, which exceeds the surface escape velocity of the host star (552 km s$^{-1}$). The single Gaussian fitting results for the He\,{\sc{i}} 6680Å contrast profiles are presented in panels D, E and F of Fig. \ref{CME_4_linefitting}, showing that the intensity and red shift of the He\,{\sc{i}} line both gradually decrease. The He\,{\sc{i}} line does not reveal two redshifted components as in the case of the H$\alpha$ line, probably due to the lower S/N. Obvious emission and small redshifts can also be seen in the Mg\,{\sc{i}} triplet and Fe\,{\sc{ii}} 5170.47Å   lines.

\subsubsection{Doppler shifts and asymmetries of the chromospheric lines in CME candidate 3}

The high-speed redshifted component in panels A, B and C of Fig. \ref{CME_4_linefitting} (green) may be caused by a CME propagating in the direction away from the Earth, while the low-speed redshifted component (pink) likely results from coronal rain falling back to the stellar surface. The central velocity of the high-speed component shows a weak increasing trend. The average central velocity is 315 km s$^{-1}$, and the largest value of the maximum velocities exceeds the surface escape velocity of the host star. Therefore, the high-speed component may be caused by a backward moving CME. A possible reason for the gradually decreasing maximum velocity is that as the CME expands during the propagation, the density of the outer region with a faster propagation speed decreases more, so its signal weakens in the line profiles. Another possible scenario is that the rapid propagation region of the CME is blocked by the star as the CME propagates away from the Earth, leaving just the slower-propagation region of the CME to be observed by the telescope. It should be noted that this may be a non-radial propagating CME occurring near the stellar limb, thus allowing both the flare and the backward propagating CME to be observed. Non-radial eruptions mean that CME motions deviate from the radial direction, often due to magnetic obstacles (coronal holes, helmet streamers, etc.) near the source regions of CMEs. Such a phenomenon is frequently observed on the Sun \citep{2013SoPh..287..391P, 2013ApJ...773..162B, 2020AdSpR..65.1654C}. The trajectories of many solar non-radial eruptions deviate from radial propagation by around 30 degrees and some even reach 90 degrees \citep{2001SoPh..203..119F, 2015SoPh..290.3343L, 2018ApJ...862...86Y}. Based on the high-speed component shown in Fig. \ref{CME_4_linefitting}A, the minimum mass of the possible CME was estimated to be $1.85 \times 10^{18}$ g. 
   
The central and maximum velocities of the low-speed component gradually decrease, and the central velocities are lower than 100 km s$^{-1}$. These characteristics are similar to those of the coronal rain on the Sun. A possible scenario is that one footpoint of the flare loop is located close to the stellar limb and the other footpoint is blocked by the stellar disk. In that case, cool coronal rain falling back to the stellar surface along the flare loop may result in a redshifted emission in the H$\alpha$. The typical speed of coronal rain is between 30 km s$^{-1}$ and 150 km s$^{-1}$ \citep{2016ApJ...818..128O, 2020PPCF...62a4016A, 2021ApJ...910...82L, 2022A&A...659A.107C}. In solar observations, redward asymmetries of the H$\alpha$ line caused by coronal rain have been identified \citep{2014SoPh..289.4117A}. \cite{2018A&A...615A..14F} analyzed the H$\alpha$ asymmetries of M dwarfs and suggested that a possible reason for the red wing enhancement is coronal rain. Moreover, if this flare occurs near the stellar limb, the decrease of the emission in the H$\alpha$ line core may be caused by the decreasing projected area of the flare ribbons as the star rotates. The observed red shifts of the Mg\,{\sc{i}} triplet, Fe\,{\sc{ii}} 5170.47Å and He I 6680Å lines may be caused by a combined effect of CME and coronal rain, which cannot be separated due to the low S/N in these spectral lines.
   
   In order to verify the activity of the host star for CME candidate 3, we collected two sets of photometric data for the host star from the TESS database \citep{2014SPIE.9143E..20R, 2015JATIS...1a4003R}. After normalizing the flux and using the power spectral density to calculate the period of starspot activity, we plot these two sets of data in panels A and B in Fig. \ref{CME_4_TESS}. Both sets of data show changes in the light curves caused by starspots, and the period of the spot activity is about 0.45 days. In addition, it can be seen from Fig. \ref{CME_4_TESS} that the host star produced multiple superflares during the two observation periods of the TESS. The light curve of the largest superflare is zoomed-in in panel C. The duration of this superflare is 58 minutes and its energy was estimated to be $4.37 \times 10^{34}$ erg though a method used by \cite{2019ApJS..243...28L}. The photometric data of the TESS suggests that the host star of CME candidate 3 is a very active star, which is likely to generate superflares and CMEs to cause the Doppler shifts and asymmetries of the spectral lines, as mentioned above. For the host stars of the other two candidates, we were unable to collect other valuable data.

\section{Summary and future perspectives}

\begin{table*}
\caption{Parameters of the three CME candidates. }             
\label{table:3}      
\centering
\begin{tabular}{cccc}     
\hline\hline       
                     
CME candidate & Mass (g) & Maximum bulk velocity (km s$^{-1}$) & Maximum velocity (km s$^{-1}$)  \\ 
\hline    
LAMOST obsid 876604049 &  $4.12 \times 10^{18}$ & --124 & --473  \\         
LAMOST obsid 635003103 &  $8.84 \times 10^{17}$ & --71  & --151   \\  
LAMOST obsid 624510064 &  $1.85 \times 10^{18}$ & 341 & 698  \\     
\hline                  
\end{tabular}
\end{table*}
   
   In this work, we searched for stellar CMEs on late-type main-sequence stars ($\rm T_{eff} < 6000$ K, $\rm log [g/(cm\ s^{-2})] > 4.0$) in the LAMOST-MRS. The search sample contains 1,379,408 LAMOST-MRS spectra, which come from 226,194 late-type main-sequence stars. By searching for the asymmetric H$\alpha$ line profiles in the sample and conducting visual inspections, we finally identified three possible stellar CMEs. The relevant parameters of the three CME candidates are summarized in Table \ref{table:3}, including the CME mass, maximum bulk velocity and maximum velocity. The host star of the three CME candidates are three M dwarfs. After performing a double Gaussian fitting for the H$\alpha$ contrast profiles of the continuously observed spectra of the three CME candidates, we obtained the Doppler shifts of the asymmetric emission components. In light of the observations and theoretical models of solar flares and CMEs, the possible physical processes responsible for the asymmetric components of the H$\alpha$ line profiles in the three CME candidates are discussed. The main results for these three stellar CME candidates are summarized as follows:

   \begin{enumerate}
      \item The host star of CME candidate 1 (LAMOST obsid 876604049) is an M1-type main-sequence star. CME candidate 1 appears to occur in the impulsive phase of the associated stellar flare, when intensities of the Mg\,{\sc{i}} triplet, Fe\,{\sc{ii}} 5170.47Å, H$\alpha$ and He\,{\sc{i}} 6680Å lines all gradually increase. The Mg\,{\sc{i}} triplet and H$\alpha$ lines exhibit obvious blue-wing enhancements, which are likely caused by a CME. The minimum mass of the CME was estimated to be $4.12 \times 10^{18}$ g. The central and maximum velocities of the blue wing component of H$\alpha$ gradually increase. The H$\alpha$ line also exhibits a redshifted component that is likely caused by the chromospheric condensation.     
 
      \item The host star of CME candidate 2 (LAMOST obsid 635003103)  is an M4-type main-sequence star. The H$\alpha$ line shows a blueshifted component that is likely caused by a CME. The minimum mass of the CME was estimated to be $8.84 \times 10^{17}$ g. An additional broad emission component of the H$\alpha$ line is probably caused by the Stark broadening effect. 
 
      \item The host star of CME candidate 3 (LAMOST obsid 624510064) is an M2-type main-sequence star. This CME candidate might occur in the decay phase of the associated stellar flare.  The H$\alpha$ line profiles show two redshifted components, one of which may be caused by a CME propagating in the direction away from the observer and its minimum mass was estimated to be $1.85 \times 10^{18}$ g. The other redshifted component may be caused by flare-triggered coronal rain. In addition, our analysis of the photometric data of the TESS shows that the host star of CME candidate 3 is a very active star, which is capable of generating superflares and CMEs to cause the asymmetries of the spectral lines.
   \end{enumerate}
   
   From our large sample, we only detected three possible stellar CMEs. The probability of detecting stellar CME is much smaller than that of superflares \citep{2012Natur.485..478M, 2016NatCo...711058K} on late-type main-sequence stars. \cite{2021A&A...646A..34K} also detected only 6 CME candidates among the 630,000 late-type main-sequence stars in SDSS. \cite{2020A&A...637A..13M} tried to search for stellar CMEs from 2000 high-dispersion spectra of AD Leo, but did not find a CME candidate with a velocity exceeding the escape velocity of the host star. \cite{2019A&A...623A..49V} found 478 spectra with line asymmetries from 25 stars in about 5500 spectra and suggested that these stars may show asymmetries in their spectral line profiles multiple times per day. The reason why \cite{2019A&A...623A..49V} detected more CME candidates may be due to selection effect. Their samples are mainly active M dwarfs. \cite{2020MNRAS.493.4570L} did not detect CME candidates in FGK-type main-sequence stars using the data sources that were partially similar to that of \cite{2019A&A...623A..49V}. Therefore, the reason why there are fewer stellar CME candidates detected from our sample may be as follows. First, although the LAMOST-MRS has observed many spectra, the time domain survey is in the early stage and the number of spectra observed for the same target is still too few. As the time-domain spectral survey continues, the LAMOST-MRS will conduct many more observations of stars and we may find more stellar CMEs in the future. Second, the torus instability of magnetic flux ropes is one of the main triggering mechanisms of solar CMEs. On some stars, the background coronal magnetic field decreases slowly with altitude, which may suppress the torus instability and thus reduce the rate of stellar CME eruptions \citep{2022MNRAS.509.5075S}.  The strong magnetic field on some stars may also suppress CME eruptions \citep{2018ApJ...862...93A, 2020ApJ...900..128L}. In one word, the number of stellar CME candidates is still very small, and long-term continuous spectroscopic observations are highly desired for expanding stellar CME samples.   

 \begin{acknowledgements}
      This work is supported by the NSFC grants 12103004, 11825301 and 11790304, the fellowship of China National Postdoctoral Program for Innovative Talents (BX2021017), and the Strategic Priority Research Program of CAS (XDA17040507). We thank the referee and Professor Shenghong Gu for the very constructive suggestions and helpful comments.  Guoshoujing Telescope (the Large Sky Area Multi-Object Fiber Spectroscopic Telescope LAMOST) is a National Major Scientific Project built by the Chinese Academy of Sciences. Funding for the project has been provided by the National Development and Reform Commission. LAMOST is operated and managed by the National Astronomical Observatories, Chinese Academy of Sciences. Part of the data in this paper was collected by the TESS mission and downloaded from the Mikulski Archive for Space Telescopes (MAST). This research has made use of the SIMBAD database, operated at CDS, Strasbourg, France. This work uses the software TOPCAT \citep{2005ASPC..347...29T}. 
\end{acknowledgements}

%
%

\bibliographystyle{aa} 
\bibliography{lhpref} 

\end{document}